\documentclass{ws-p10x7}
\newcommand{\beq}{\begin{equation}}
\newcommand{\eeq}{\end{equation}}
\newcommand{\bea}{\begin{eqnarray}}
\newcommand{\eea}{\end{eqnarray}}
\def\al{\alpha}
\def\be{\beta}
\def\ga{\gamma}
\def\de{\delta}
\def\si{\sigma}

\def\O{{\cal{O}}}
\def\wt{\widetilde}
\def\ol{\overline}
\def\l{\left}
\def\r{\right}

\begin{document}

\title{Fully  supersymmetric CP violations \\ in the kaon system }



\author{S. Baek,~J.-H. Jang, ~~P. Ko$^*$ ~~and ~~J. H. Park} 
\address{Dep. of Physics, KAIST, Taejon 305-701, KOREA
\\ E-mail: $*$pko@charm.kaist.ac.kr}

\twocolumn[\maketitle\abstract{
We show that, on the contrary to the usual claims, fully supersymmetric 
CP violations in the kaon system are possible through the gluino 
mediated flavor changing interactions. Both $\epsilon_K$ and ${\rm Re} 
(\epsilon' / \epsilon_K)$ can be accommodated for relatively 
large $\tan\beta$ without any fine tunings or contradictions 
to the FCNC and EDM constraints. 
}]




Recent observation of ${\rm Re} (\epsilon' / \epsilon_K )$ by KTeV 
collaboration, 
$
{\rm Re} ( \epsilon' / \epsilon_K ) = (28 \pm 4) \times 10^{-4}
$
\cite{ktev}, 
nicely confirms the earlier NA31 experiment \cite{na31}
$
{\rm Re} ( \epsilon' / \epsilon_K ) = (23 \pm 7) \times 10^{-4}.
$
This nonvanishing number indicates unambiguously the existence of  CP 
violation in the decay amplitude $(\Delta S = 1)$. Along with another 
CP violating parameter known for long time,  
$\epsilon_K = e^{i \pi/4} ~( 2.280 \pm 0.013 ) \times 10^{-3}$ \cite{pdg},
these two parameters quantifying CP violations in the kaon system can be 
accommodated by the KM phase in the 
standard model (SM). 
The SM prediction for ${\rm Re} ( \epsilon' / \epsilon_K )  $ 
is about $5 \times 10^{-4}$ and lies in the lower side of the data, 
although theoretical uncertainties 
are rather large \cite{buras_1}.  

However, it would be interesting to consider a possibility that 
both $\epsilon_K$ and ${\rm Re} ( \epsilon' / \epsilon_K )  $
have their origins entirely different from the KM phase 
in the SM, in particular in supersymmetric models.   
In this talk, we argue that all the observed CP violating phenomena in the 
kaon system
in fact can be accommodated in terms of a {\it single} complex number 
$( \delta_{12}^d )_{LL}$ that parameterizes the squark mass mixings in the 
chirality and flavor spaces 
for relatively large $\tan\beta$ without any fine tuning or any 
contradictions with experimental data on FCNC, even if we assume
 $\delta_{\rm KM}=0$ as in this talk.
This talk is based on Ref.~\cite{ours}.

In order to study the gluino  mediated flavor changing phenomena 
in the quark sector,
it is convenient to use the so-called mass insertion approximation (MIA) 
\cite{mia}.
The parameters $( \delta_{ij}^d )_{AB}$ characterize the size of the 
gluino-mediated flavor ($i,j$) and chirality ($A,B$) changing amplitudes. 
They may be also CP violating  complex numbers. 

Now, if one saturates $\Delta m_K$ and $\epsilon_K$ with 
$( \delta_{12}^d )_{LL}$ alone, the resulting ${\rm Re} ( \epsilon^{'} / 
\epsilon_K )$ is too small by more than an order of magnitude,
unless one invokes some finetuning \cite{gabriel}.   
On the other hand, if one saturates ${\rm Re} ( \epsilon^{'} / \epsilon_K )$ 
by $| {\rm Im} ( \delta_{12}^d )_{LR} | \sim 10^{-5}$, the resulting 
$\epsilon_K$ is too small by more than an order of magnitude, unless one 
invokes some finetuning.   
Therefore the folklore was that the supersymmetric contributions to 
${\rm Re} ( \epsilon^{'} / \epsilon_K )$ is small.  Recently, Masiero and 
Murayama showed that this conclusion can be evaded in generalized SUSY models 
\cite{mm} with a few reasonable assumptions on the size of the 
$(\delta_{12}^d)_{LR}$.  
But in their model, one has to introduce a new 
CP violating parameter $( \delta_{12}^d )_{LL} $  
in order to generate $\epsilon_K$ and also
predict too large neutron EDM which is very close to the current upper limit.

In the following, we show that there is another generic way to evade 
this folklore in supersymmetric models if $|\mu \tan\beta|$ is relatively 
large, say  $\sim 10 - 20$ TeV.  Moreover, both $\epsilon_K$ and 
${\rm Re} ( \epsilon^{'} / \epsilon_K )$ can be generated by a single 
CP violating complex parameter in the MSSM. In other words, fully 
supersymmetric CP violations are possible in the kaon system. 
The argument goes as follows : 
if $| ( \delta_{12}^d )_{LL} | \sim O(10^{-3} - 10^{-2})$ with the phase 
$\sim O(1)$ saturates $\epsilon_K$, this same parameter can
lead to a sizable ${\rm Re} (\epsilon' / \epsilon_K )$ through the 
$( \delta_{12}^d )_{LL}$ insertion followed by the FP $(LR)$ mass 
insertion, which is proportional to 
\[
(\delta_{22}^d)_{LR} \equiv m_s ( A_s^* - \mu \tan\beta) / \tilde{m}^2 
\sim O(10^{-2}), 
\]
where $\tilde{m} $ denotes the common squark mass in the MIA.
It should be emphasized that the induced $( \delta_{12}^d )_{LR}^{\rm ind}
\equiv ( \delta_{12}^d )_{LL} \times (\delta_{22}^d)_{LR} $ 
is different from the conventional  $ ( \delta_{12}^d )_{LR}$ in the 
literature. 
The $LR$ mixing $( \delta_{12} )_{LR}^{\rm ind}$ 
induced by $( \delta_{12}^d )_{LL}$  is typically very small in size 
$\sim O(10^{-5})$, but this is enough to generate the full size of 
${\rm Re} (\epsilon' / \epsilon_K)$ as shown below. 
Thus the usual folklore can be simply evaded.
Our spirit to generate supersymmetric ${\rm Re} 
(\epsilon^{'} / \epsilon_K)$ is different from  Ref.~\cite{mm}, 
where the $LR$ mass matrix form is assumed to be similar to
the Yukawa matrix so that they predict the neutron EDM to be close to the 
current upper limit. On the other hand, we do not assume any specific 
flavor structure in trilinear $A$ couplings. 
Also our model does not suffer from the
EDM constraint at all.

Let us first consider the gluino-squark contributions to the 
$K^0 - \overline{K^0}$ mixing due to two insertions of 
$( \delta_{12}^d )_{LL}$. The corresponding $\Delta S = 2$ effective 
Hamiltonian is given by 
\[
{\cal H}_{\rm eff} (\Delta S =2) 
= C_{1} \ol{d}_L^\al \ga_\mu s_L^\al ~\ol{d}_L^\be \ga^\mu s_L^\be
\]
with the Wilson coefficient $C_{1}$ being 
\bea
  C_1 = -{\al_s^2 \over 216 \wt{m}^2} 
         (\de^d_{12})^2_{LL} ~ f_1 (x) 
\eea
Here, $x=m^2_{\wt{g}}/\tilde{m}^2$ and the loop function $f_1 (x)$ 
is given in~\cite{mia}.

\begin{figure}[h]
\centerline{\epsfxsize=5.0cm \epsfbox{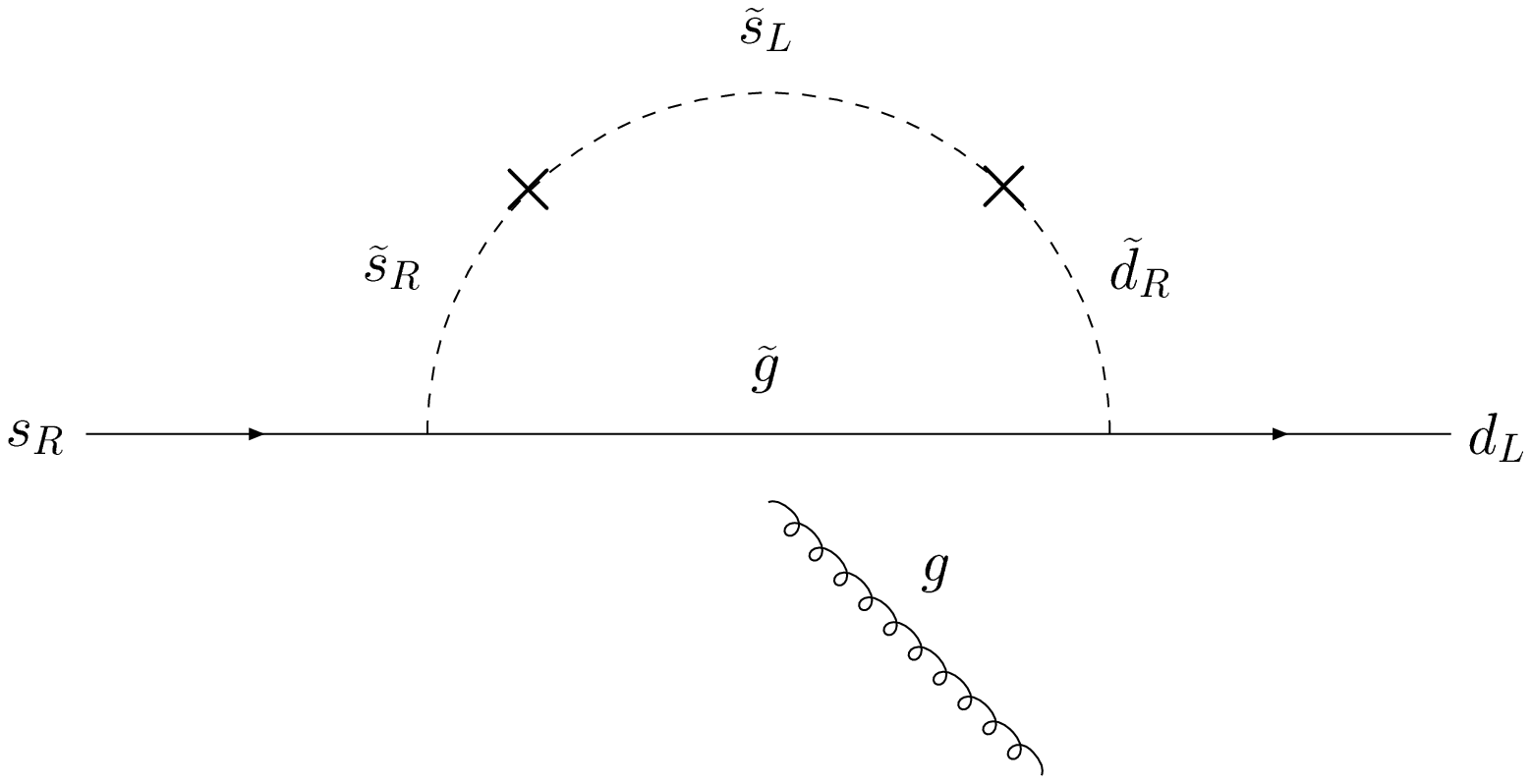}}
\vspace{.5cm}
\caption{ Feynman diagram for $\Delta S=1$ process. The cross denotes the 
flavor changing $(LL)$ and the flavor preserving $(LR)$ mixings, 
respectively.}
\label{fig1}
\end{figure}

Now we turn to the $\Delta S=1$ effective Hamiltonian 
${\cal H}_{\rm eff} ( \Delta S = 1 ) = \sum_{i=3}^8 C_i \O_i$.
The $sdg$ operator $O_8$ which is relevant to 
${\rm Re} (\epsilon^{'}/\epsilon_K )$ is defined as 
\begin{equation}
\O_8 = {g_s \over 4 \pi}  m_s \ol{d}_L^\al \si^{\mu\nu} T^a s_R^\al 
                G^a_{\mu\nu}, 
\end{equation}
and other four quark operators $O_{i=3,...,6}$ and the corresponding Wilson 
coefficients from $C_3$ to $C_8$ with a single mass
insertion are available in the literature~\cite{mia}. 
If we consider the penguin diagram Fig.~\ref{fig1} with the double mass 
insertion, the Wilson coefficient $C_8$ is given by 
\begin{equation}
\label{newdel1}
   C_8^{(2)} = {\al_s \over \wt{m}^2} \; {m_{\wt{g}} \over m_s} 
           (\de_{12}^d)_{LR}^{\rm ind} ~
M_8 (x) 
\end{equation}
where the explicit form of the $\sim O(1)$ function 
$M_8 (x)$ can be found in Ref.~\cite{ours}. 
Since $C_8^{(2)}$ is proportional to 
$m_{\tilde{g}}/ m_s$, it is very important for generating
${\rm Re} (\epsilon^{'} / \epsilon_K)$ even if 
$(\delta_{12}^d )_{LR}^{\rm ind}$ is fairly small.
%
%

It is straightforward to calculate $\epsilon_K$ and 
$\epsilon^{'} / \epsilon_K$ using the same parameters as in Ref.~\cite{buras}
with $m_s (2~ {\rm GeV}) = 130~ {\rm MeV}$. 
The corresponding SM prediction for ${\rm Re} ( \epsilon^{'} / \epsilon_K )
= 5.7 \times 10^{-4}$.  
For those points which satisfy 
$\Delta m_K ( {\rm SUSY} ) \leq \Delta m_{K} ({\rm exp})$ 
and $| \epsilon_K ({\rm SUSY}) - \epsilon_K ({\rm exp}) | < 1 \sigma $, 
we plot $\epsilon' / \epsilon_K$ in Fig.~2 
as functions of the modulus $r$  and the phase $\varphi$ 
of the parameter $(\delta_{12}^d)_{LL} \equiv r e^{i \varphi}$  
for the common squark mass $\tilde{m}= 500$ GeV. 
The upper (lower) rows correspond to 
$\wt{A_s} \equiv (A_s - \mu^* \tan\beta ) = -10 (20)$ TeV. 
It is clear that both $\epsilon_K$ and 
${\rm Re} (\epsilon' / \epsilon_K)$ can be nicely 
accommodated with a single complex number $(\delta_{12}^{d})_{LL}$ with 
$\sim O(1)$ phase in our 
model without any difficulty, if $| \mu| $ and $\tan\beta$ is relatively 
large so that $|\wt{A_s}| $ becomes a few tens of TeV.  
\begin{figure}[h]
\centerline{\epsfxsize=8.0cm \epsfbox{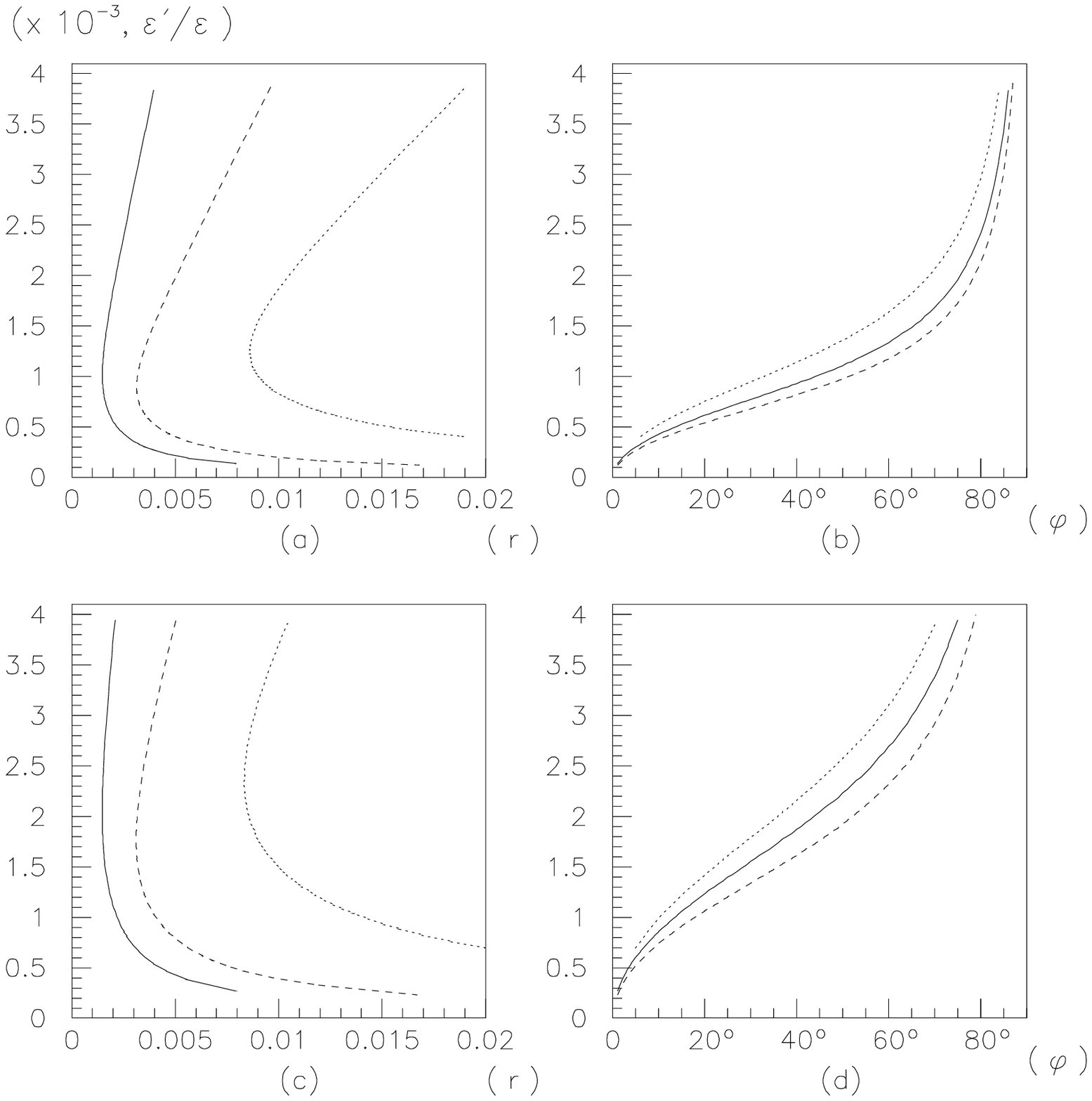}}
\caption{ $Re ( \epsilon' / \epsilon_K )$ as a function 
of the modulus $r$ [(a) and (c)] and 
the phase $\varphi$ [(b) and (d)] of the parameter $(\delta_{LL}^d )_{12}$ 
with $\wt{A_S}$ to be $-10~TeV$ ((a),(b)) and 
$-20~TeV$ ((c),(d)).  
The common squark mass is chosen to be $\wt{m} = 500$ GeV, and the solid, 
the  dashed and the dotted curves correspond to $x = 0.3, ~1.0, ~2.0$, 
respectively.
}
\label{fig2}
\end{figure}
It is important to realize that in our model there is no conflict with 
neutron EDM from $( \delta_{12}^d )_{LL}$, since the EDM is generated by  
another parameter $\l(\de^d_{11} \r)_{LR}$, which can be taken as real 
independent of $( \delta_{12}^d )_{LL}$.  
This is in sharp contrast with the case of Ref.~\cite{mm} where the 
neutron edm is inevitably close to the current upper limit.
Note that we are not assuming any 
specific flavor structures in the $A$ terms at all, unlike many other recent 
models in the literatures. 

In conclusion, we showed that both $\epsilon_K$ and 
${\rm Re} (\epsilon' / \epsilon_K )$ 
can be accommodated with a single CP violating and flavor changing 
down-squark mass matrix elements $[ (\delta_{12}^d )_{LL} \sim 10^{-3} ]$ 
without any fine tuning or any conflict with the data on FCNC processes,
if $| \mu \tan\beta| \sim O(10)$ TeV.
Our mechanism utilizes this FC $LL$ mass insertion along with the FP $LR$ 
mass insertion propotional to $(\delta_{22}^d )_{LR} \sim 10^{-2}$.  
The latter is {\it generically present in any SUSY models} including the MSSM, 
and thus there is no fine tuning in our model for accommodating both 
$\epsilon_K$ and ${\rm Re} (\epsilon' / \epsilon_K)$ in terms of a single
$(\delta_{12}^d )_{LL} $. Phrasing differently, {\it the SUSY $\epsilon_K$ 
problem implies SUSY $\epsilon^{'}$ problem if 
$| \mu \tan \beta | ( \sim O(10-20)$  TeV) is relatively large}.  
One can also consider our mechanism in the more minimal SUSY model, where 
$(\delta_{22})^d_{LR}$ is proportional to $m_b ( A_b - \mu \tan\beta)$ 
so that $A_b - \mu \tan\beta$ may be lowered significantly.
The best discriminant between our model and the SM model would be probably
the branching ratios for $K\rightarrow \pi \nu\nu$ and CP violations in 
$B$ decays.  
For example, the time dependent asymmetry in $B^0 \rightarrow J/\psi K_S$,
$\sin 2\beta$, can be completely different from the SM prediction. 
Also if the KM phase is nonzero, there will 
be additional constributions to $\epsilon_K$ and ${\rm Re} (\epsilon^{'} / 
\epsilon_K)$ from the  SM and other SUSY loop diagrams.
All these finer details 
will be discussed elsewhere in the forthcoming publication \cite{future}. 


 

%

 
 
\section*{Acknowledgments}
This work is supported by BK21 program of Ministry of Education.


\end{document}